\documentclass[useAMS,usenatbib]{mn2e}
\usepackage{aas_macros,graphicx,times,multirow,amsmath,bbold}
\usepackage[]{color,hyperref}

\title[Polarization of neutron star surface emission]{Polarization of neutron star surface emission: a systematic analysis}
\author[R. Taverna et al.]{R.~Taverna$^{1}$\thanks{E-mail:
\href{mailto:taverna@pd.infn.it}{taverna@pd.infn.it}},
R.~Turolla$^{1,\, 2}$, D.~Gonzalez Caniulef$^{2}$, S.~Zane$^{2}$, F.~Muleri$^{3}$, P.~Soffitta$^{3}$
\smallskip\\
$^1$Department of Physics and Astronomy, University of Padova, via Marzolo 8, 35131 Padova, Italy\\
$^2$Mullard Space Science Laboratory, University College London, Holmbury St. Mary, Surrey, RH5 6NT, UK\\
$^3$INAF-IASF Roma, Via del Fosso del Cavaliere 100, 00133 Roma, Italy\\
}

\date{Accepted \ldots. Received \ldots; in
original form \ldots} \pagerange{\pageref{firstpage}--\pageref{lastpage}} \pubyear{2015}

\begin{document}

\label{firstpage}
\maketitle
\begin{abstract}

New-generation X-ray polarimeters currently under development
promise to open a new window in the study of high-energy
astrophysical sources. Among them, neutron stars appear
particularly suited for polarization measurements. Radiation from
the (cooling) surface of a neutron star is expected to exhibit a
large intrinsic polarization degree due to the star strong
magnetic field ($\approx 10^{12}$--$10^{15}$ G), which influences
the plasma opacity in the outermost stellar layers. The
polarization fraction and polarization angle as measured by an
instrument, however, do not necessary coincide with the intrinsic
ones derived from models of surface emission. This is due to the
effects of quantum electrodynamics in the highly magnetized vacuum
around the star (the vacuum polarization) coupled with the
rotation of the Stokes parameters in the plane perpendicular to
the line of sight induced by the non-uniform magnetic field.
Here we revisit the problem and present an efficient
method for computing the observed polarization fraction and
polarization angle in the case of radiation coming from the entire surface of a neutron
star, accounting for both vacuum polarization and geometrical
effects due to the extended emitting region. 
Our approach is fairly general and is illustrated in the
case of blackbody emission from a neutron star with either a
dipolar or a (globally) twisted magnetic field.

\end{abstract}

\section{Introduction}
\label{intro}

Polarization measurements of radiation coming from astrophysical
sources helped in improving our knowledge about the
physical and geometrical properties of a variety of systems, from
black holes to gamma-ray bursts \cite[e.g.][for a review]{trippe14}. In this respect, neutron stars (NSs) are among the
most promising targets for polarimetry due to their strong
magnetic field which is expected to induce a large degree of
polarization of the emitted radiation.

Radio and optical polarimetry has been already used to derive the
orientation of the magnetic and rotation axes of radio pulsars
(\citealt{ManchesterTaylor,LyneManchester}; see also
\citealt{PavlovZavlin}). The discovery over the last two decades
of new classes of X-ray bright, radio-silent NSs with very faint
(if any) optical counterparts (chiefly the magnetar candidates,
e.g. \citealt{mereg08,turolla15}, and the X-ray Dim Isolated
Neutron Stars, XDINSs, e.g. \citealt{turolla09,kaspi10}) renewed
the interest in possible polarization measurements at X-ray
energies in NS sources. Despite some efforts were made in the past
to measure polarization in the X-rays, mainly with the OSO-8 and
INTEGRAL satellites
(\citealt{Weisskopfetal78,Hughesetal,Deanetal}; see also
\citealt{Kislatetal}), the poor sensitivity of past
instrumentation did not lead to conclusive results. A new window
opened in the last years, with the advent of new-generation X-ray
polarimeters, like XIPE
\footnote{http://www.isdc.unige.ch/xipe}, IXPE and PRAXyS
\footnote{\citet{Weisskopfetal}, \citet{Jahodaetal}} (recently selected for the study
phase of the ESA M4 and NASA SMEX programmes respectively), which are based on the photoelectric
effect and provide a dramatic increase in sensitivity over an
energy range $\sim 1$--30 keV \cite[see][]{Bellazzinietal}.
X-ray polarimeters derive polarization observables by detecting a
modulation in the azimuthal distribution of events in the focal
plane. Actually, while a measure of the circular polarization
degree is possible in the optical band \cite[see
e.g.][]{Wiktorowiczetal}, current instruments, based on the
photoelectric effect or Compton scattering, can only provide
information about linear polarization \cite[][]{FabianiMuleri}.

From a theoretical viewpoint, polarization observables (the polarization fraction and the polarization angle) are
conveniently expressed through the Stokes parameters. The
comparison between the polarization properties of the photons
emitted at the source and those measured at earth is not
straightforward for two main reasons. The first is that the Stokes
parameters are defined with respect to a given frame, which is in
general different for each photon. When the Stokes parameters
relative to the different photons are added together, care must be
taken to rotate them, so that they are referred to the same frame, which coincides with
the frame in the focal plane of the detector. This effect becomes important 
every time radiation comes from a spatial region endowed with a non-constant 
magnetic field, and will be referred to as ``geometrical effect'' in the
following. The second issue,
which typically arises in NSs, is related to ``vacuum
polarization''. 
In the presence of a strong magnetic field, quantum electrodynamics (QED)
alters the dielectric and magnetic properties of the vacuum outside 
the star, substantially affecting polarization \cite[][]{HeylShaviv2002}.
Because of this, (100\% linearly polarized) photons emitted by the surface 
will keep their polarization state up to some distance from the star, as 
they propagate adiabatically. This implies that the degree of polarization 
and the polarization angle, as measured at infinity, depend also on the 
extension of the ``adiabatic region'', which in turn depends on the photon energy
and on the magnetic field.

The observed polarization properties of radiation from isolated NSs were
investigated in the past both in connection with the emission from
the cooling star surface and the reprocessing of photons by
magnetospheric electrons through resonant Compton scattering, a
mechanism which is thought to operate in magnetars. 
\citet{PavlovZavlin} studied the case of thermal emission from the entire
surface of a NS covered by an atmosphere, without accounting for QED and
geometrical effects. A quite complete analysis of the observed polarization 
properties of surface emission from a neutron star has been presented by 
\citet{Heyletal2003}, while \citet{LaiHo2003} and \citet{vanAdelsbergPerna} 
focused on the role played by the vacuum resonance\footnote{A Mikheyev-Smirnov-Wolfenstein 
resonance which may induce mode conversion in X-ray photons for typical magnetar-like fields
($B\ga 10^{14}$ G).}, which occurs
in the dense atmospheric layers, on the polarization, and may provide a 
direct observational signature of vacuum polarization. The two latter
works were restricted to the case of emission from a small hot spot on
the NS surface, over which the magnetic field can be treated as
uniform, therefore no account for rotation of the Stokes parameters was required.
\citet{Fernandez+Davis} and \citet{Tavernaetal} have shown that 
X-ray polarization measurements can provide independent estimates of the 
geometrical and physical parameters and probe QED effects in the strong 
field limit in magnetar sources.

In this paper we re-examine the problem and present a
simplified, efficient method to derive the observed polarization
properties of radiation emitted from the entire surface of a NS.
Our results are in agreement with those of \citet{Heyletal2003}
and \citet{vanAdelsbergPerna}, and our faster approach allows to
systematically explore the dependence of the polarization observables
on the different geometrical and physical quantities. 
In particular, we discuss the difference between the polarization 
properties of the radiation emitted by the star and those measured 
at earth , which is induced by geometrical and QED effects. This aspect, 
which is crucial when one needs to reconstruct the star properties from 
the observed quantities, has not been systematically investigated in previous 
works. A complete study based on physically consistent models of surface 
emission is outside the scope of this analysis, and we just assume a simple 
model in which the surface emission is a (isotropic) blackbody and the magnetic 
field is dipolar (or a globally twisted dipole field). The outline of the 
paper is as follows. The theoretical framework is introduced in section 
\ref{sec:theoreticaloverview}. In section \ref{sec:alphadistribution} calculations 
and results are presented, while section \ref{sec:conclusion} contains a
discussion about our findings and the conclusions.

\section{Theoretical overview} \label{sec:theoreticaloverview}

In this section we briefly summarize some basic results about the
evolution of the polarization state of electromagnetic radiation
propagating in a strongly magnetized vacuum. Although the
considerations we present below are focused on radiation travelling in the
surroundings of a neutron star, they hold quite in general.

\subsection{Photon polarization in strong magnetic fields}
\label{subsec:vacuumpolarization}

In the presence of strong magnetic fields
photons are linearly polarized in two normal modes: the ordinary mode (O-mode),
in which the electric field oscillates in the plane of the propagation vector
$\boldsymbol{k}$ and the local magnetic field $\boldsymbol{B}$, and the
extraordinary mode (X-mode), in which, instead, the electric field
oscillates perpendicularly to both $\boldsymbol{k}$ and
$\boldsymbol{B}$. This holds for  photon energies below the electron cyclotron energy  \cite[$E< E_{\mathrm{c}e} = \hbar eB/m_ec
\simeq 11.6(B/10^{12}\,\mathrm{G})$ keV;][]{GnedinPavlov}, which implies $B\ga 10^{11}$ G at X-ray energies, whereas $B$ can be as low as
$\sim 10^{10}$ G in the optical band.
Moreover, the polarization state of
photons propagating in vacuo is also influenced by the effects of vacuum
polarization (\citealt{HeylShaviv2000}, \citeyear{HeylShaviv2002}, \citealt{harlai06}).
According to QED, in fact, photons can temporarily convert
into virtual $e^{\pm}$ pairs. The strong magnetic field polarizes the pairs, modifying the dielectric, $\boldsymbol{\varepsilon}$,
and magnetic permeability, $\boldsymbol{\mu}$, tensors of the
vacuum, which would coincide with the unit tensor otherwise.

Fixing a reference frame $(x,\, y,\, z)$ with the $z$-axis along 
the photon propagation direction $\boldsymbol{k}$, and the $x$-axis perpendicular 
to both $\boldsymbol{k}$ and the local magnetic field $\boldsymbol{B}$, 
the evolution of the wave electric field is governed by the following 
system of differential equations (see \citealt{Fernandez+Davis,Tavernaetal}; see also \citealt{HeylShaviv2002} 
for a different, albeit equivalent, formulation)

\begin{equation} \label{eqn:odevacuumpol}
\arraycolsep=1.4pt\def\arraystretch{2.2}
\begin{array}{ccc}
\dfrac{\mathrm{d}A_x}{\mathrm{d}z} &=& \dfrac{\mathrm{i}k_0\delta}{2}
(MA_x+PA_y) \\
\dfrac{\mathrm{d}A_y}{\mathrm{d}z} &=& \dfrac{\mathrm{i}k_0\delta}{2}
(PA_x+NA_y)\,.
\end{array}
\end{equation}
Here $\boldsymbol{A}=(A_x,\,A_y)=(a_x\mathrm{e}^{-\mathrm{i}\varphi_x},\,
a_y\mathrm{e}^{-\mathrm{i}\varphi_y})$ is the electric field
complex amplitude, $k_0=\omega/c$ with $\omega$ the photon angular
frequency and the adimensional quantities $\delta$, $M$, $N$ and
$P$ depend on the (local) magnetic field strength; in particular
it is \mbox{$\delta=(\alpha_\mathrm{F}/45\pi) (B/B_\mathrm
Q)^{2}$}, where $\alpha_\mathrm F$ is the fine structure constant
and $B_\mathrm Q=4.4\times 10^{13}\ \mathrm G$ is the critical
magnetic field. As equations (\ref{eqn:odevacuumpol}) show, vacuum polarization 
induces a change in the electric field as the wave propagates: the typical
lengthscale over which this occurs is $\ell_\mathrm{A} =
2/k_0\delta \simeq 100 (B/10^{11}\, \mathrm G)^{-2}(E/1 \,
\mathrm{keV})^{-1}\, \mathrm{cm}$, where $E=\hbar\omega$. At the
same time, the magnetic field changes along the photon trajectory,
this time over a lengthscale  $\ell_\mathrm{B}=
B/|\boldsymbol{\hat{k}}\cdot\nabla B|\sim r$, where $r$ is the
radial distance. Near to the star surface it is
$\ell_\mathrm{A}\ll\ell_\mathrm{B}$ and the direction along which
the wave electric field oscillates can instantaneously adapts to
the variation of the local magnetic field direction, maintaining
the original polarization state. In these conditions, the
photon is said to propagate adiabatically and in the following we
will refer to the region in which this occurs as the adiabatic
region. However, as the photon moves outwards the  magnetic field
strength decreases ($B\propto r^{-3}\sqrt{1+3\cos^2\theta}$ for a
dipole field, where $\theta$ is the magnetic colatitude) and
$\ell_\mathrm{A}$ increases. Since $\ell_\mathrm B$ grows more
slowly, there is an intermediate region in which the
wave electric field can not promptly follow the variation of the
magnetic field any more. Finally, in the external region,
where $\ell_\mathrm{A}\gg\ell_\mathrm{B}$, the electric field
direction freezes, and the polarization modes change as the
magnetic field direction varies along the photon trajectory.

\begin{figure}
\begin{center}
\includegraphics[width=8cm]{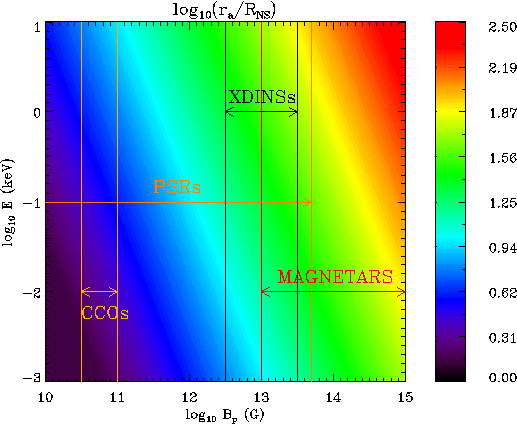}
\caption{Contour plot showing the adiabatic radius $r_\mathrm{a}$
(in units of the stellar radius $R_\mathrm{NS}$) as a function of
the polar magnetic field strength and the photon energy. The
typical $B_\mathrm P$-ranges for different classes of neutron
stars, the magnetars, the isolated, thermally emitting NSs
(XDINSs), the radio-pulsars (PSRs) and the central compact objects
(CCOs), are also shown.}
\label{fig:ra}
\end{center}
\end{figure}

The evolution of the polarization modes should be calculated
integrating equations (\ref{eqn:odevacuumpol}) from the
surface up to infinity (or, at least, up to a distance
sufficiently large to consider the complex amplitude components
$A_x$ and $A_y$  as constants). This has been the approch followed
by \citet[see also \citealt{Fernandez+Davis,Tavernaetal}]{Heyletal2003}.
However, this method requires quite long computational times, since numerical 
integration must be carried on along each ray and it is not particularly suited
for a systematic study of how the polarization observables depend on the various
physical and geometrical parameters. Since the latter is the main goal of
the present work, we resort to a simpler, approximated treatment in which 
only the adiabatic region and the external one
are included, and they are divided by a sharp edge. To this end we
introduce the adiabatic radius\footnote{This same quantity is
called the polarization-limiting radius, $r_{\mathrm{pl}}$, in previous literature
\cite[see][]{HeylShaviv2002}.} $r_\mathrm{a}$,
defined implicitly by the condition $\ell_\mathrm A = \ell_\mathrm
B$. Assuming a dipole field and purely radial photon trajectories,
it is $\ell_\mathrm B=r/3$ and hence $\ell_\mathrm A(r_\mathrm a)
= r_\mathrm a/3$. Recalling the expression for $\ell_\mathrm A$,
it follows that $r_\mathrm a/R_\mathrm{NS} \simeq 3.9\times
10^{-4}(E/1\ \mathrm{keV})^{-1}(B_\mathrm P/10^{11}\ \mathrm
G)^{-2} (R_\mathrm{NS}/r_\mathrm a)^{-6}$ and finally

\begin{equation}\label{eqn:ra}
r_\mathrm{a} \simeq 4.8\left(\frac{B_\mathrm{P}}{10^{11}
\mathrm{\,G}}\right)^{2/5}\left(\frac{E}{1\mathrm{\,keV}}\right)^{1/5}{R_\mathrm{NS}}\,,
\end{equation}
where $R_\mathrm{NS}$ is the stellar radius, $B_\mathrm P$ is the polar
strength of the dipole and $\cos\theta \sim 1$ was assumed. The adiabatic radius
depends on both the photon energy and the star magnetic field: it is
larger for stars with stronger magnetic field, and, at fixed
$B_\mathrm{P}$, it becomes smaller for less energetic photons, as shown
in Figure \ref{fig:ra}.

\subsection{Polarized radiative transfer}
\label{subsec:stokesparameters}

A convenient way to describe the polarization properties of
the radiation emitted by a source is through the Stokes parameters.
With reference to the frame $(x,y,z)$ introduced
in Section \ref{subsec:vacuumpolarization}, they are related to
the complex components of the wave electric field by
\begin{equation} \label{eqn:Stokesparameters}
\arraycolsep=1.4pt\def\arraystretch{1.8}
\begin{array}{lll}
\mathcal{I} &=& A_xA_x^* + A_yA_y^* = a_x^2 + a_y^2 \\
\mathcal{Q} &=& A_xA_x^* - A_yA_y^* = a_x^2 - a_y^2 \\
\mathcal{U} &=& A_xA_y^* + A_yA_x^* = 2a_xa_y\cos(\varphi_x-\varphi_y)\\
\mathcal{V} &=& \mathrm{i}\left(A_xA_y^* - A_yA_x^*\right) =
2a_xa_y\sin(\varphi_x-\varphi_y)\,.
\end{array}
\end{equation}
In the previous equations a star denotes the complex conjugate,
$\mathcal{I}$ is the total intensity of the wave associated to the
photon, $\mathcal{Q}$ and $\mathcal{U}$ describe the linear
polarization and $\mathcal{V}$ the circular polarization. The four
Stokes parameters satisfy the general relation $\mathcal{I}^2\geq
\mathcal{Q}^2+\mathcal{U}^2+ \mathcal{V}^2$, the equality holding
for 100\% polarized radiation. With our current choice of the
reference frame, it is $a_y=0$ for an X-mode photon and $a_x=0$
for an O-mode photon. Normalizing the Stokes parameters defined
above to the intensity $\mathcal{I}$, we can associate to an
extraordinary/ordinary photon the vectors
\begin{equation} \label{eqn:QUbasex}
\begin{array}{ccc}
\arraycolsep=1.4pt\def\arraystretch{1.1}
\left(\begin{array}{l}
\bar{\mathcal{Q}} \\
\bar{\mathcal{U}} \\
\bar{\mathcal{V}}
\end{array}\right)_\mathrm{X} = \left(\begin{array}{c}
1 \\
0 \\
0
\end{array}\right) & \ \ \ &
\arraycolsep=1.4pt\def\arraystretch{1.1}
\left(\begin{array}{l}
\bar{\mathcal{Q}} \\
\bar{\mathcal{U}} \\
\bar{\mathcal{V}}
\end{array}\right)_\mathrm{O} = \left(\begin{array}{c}
-1 \\
0 \\
0
\end{array}\right)
\end{array}\,,
\end{equation}
where a bar denotes the normalized Stokes parameters.
The evolution of the Stokes parameters mirrors that of the complex components
of the electric field given in equations (\ref{eqn:odevacuumpol})
\cite[see e.g.][]{Tavernaetal}.
Actually, in our hypothesis of $100\%$ linearly polarized thermal radiation, the
Stokes parameter $\mathcal{V}$ is always zero inside the adiabatic region. 
This implies that a circular polarization degree can arise only
as a consequence of the polarization mode evolution in the transition 
between the adiabatic and the external region. 
However, since we do not integrate equations (\ref{eqn:odevacuumpol}) in our model
we will not discuss $\mathcal{V}$ further. We verified that, even
accounting for the Stokes parameter evolution, as
obtained solving equations (\ref{eqn:odevacuumpol}) across the entire region, 
the resulting circular polarization fraction is very small at optical energies and
reaches at most a few percent in the X-ray band.

\begin{figure}
\begin{center}
\includegraphics[width=7.5cm]{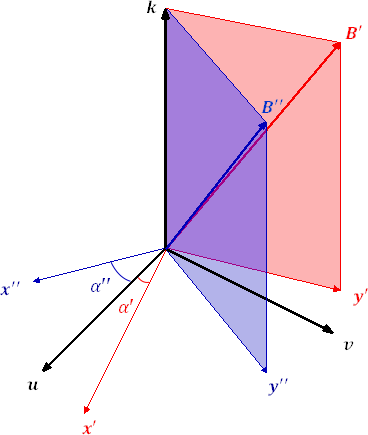}
\caption{Graphical visualization of the different reference frames
introduced in the text. $\boldsymbol{k}$ is the common direction
of propagation (LOS); $(\boldsymbol{u}$, $\boldsymbol{v})$ are the
fixed, mutually orthogonal axes of the polarimeter;
$(\boldsymbol{x}',\boldsymbol{y}')$ and
$(\boldsymbol{x}'',\boldsymbol{y}'')$ are mutually orthogonal axes
of two reference frames relative to photons coming from points
characterized by the directions $\boldsymbol{B}'$ and
$\boldsymbol{B}''$ of the  magnetic field. The angles $\alpha'$,
$\alpha''$ are also indicated.} \label{fig:uxbases}
\end{center}
\end{figure}

In order to measure the polarization properties of a given source,
a polarimeter will collect a large number of photons, each characterized by
its own set of Stokes parameters. The convenience of using the
Stokes parameters lies precisely in the fact that they are
additive \cite[e.g.][]{RybickiLightman}: the Stokes parameters associated to the whole collected
radiation (i.e. the superposition of all the received photons) are
equal to the sum of the Stokes parameters of the single photons.
However, care must be taken since the quantities in equations
(\ref{eqn:Stokesparameters}) are defined with respect to a precise
reference frame, $(x,y,z)$, that depends on the direction of the
local magnetic field $\boldsymbol{B}$ (see \S
\ref{subsec:vacuumpolarization}). As the magnetic field is in
general non-uniform across the emission region, 
its direction at a given point will depend
on the source magnetic topology. Since the
direction of the electric field of each photon varies very quickly
inside the adiabatic region, but it is frozen outside, what
actually matters is not the $B$-field direction at the
original emission point, but that at the point where the photon crosses
the adiabatic boundary $r_\mathrm{a}$, as pointed out by \citet{LaiHo2003}.
Let us call $(x_\mathrm{i},y_\mathrm{i},z_\mathrm{i})$ 
the reference frame in which the Stokes parameters associated to 
the generic photon are defined at the adiabatic radius. While the
$z_\mathrm{i}$ axes are all along the same direction
(that conicides with the observer line-of-sight, LOS),
the $x_\mathrm{i}$ and
$y_\mathrm{i}$ axes will point, in general, in different
directions for each photon (see Figure \ref{fig:uxbases}).
To sum correctly the Stokes parameters it is necessary to refer
them to the same, fixed frame, say $(u,v,w)$. This frame can be chosen
in such a way to coincide with that of the polarimeter, with $u$ and
$v$ in the detector plane and $w$ along the LOS.

Each reference frame $(x_\mathrm{i},y_\mathrm{i},z_\mathrm{i})$ is
rotated with respect to the fixed, $(u,v,w)$, frame by an angle
$\alpha_\mathrm i$ around the common $z_\mathrm i\equiv w$ axis, where
$\cos\alpha_\mathrm i= \mathbf u\cdot \mathbf x_\mathrm i$. Under
a rotation of the reference frame by an angle $\alpha_\mathrm{i}$,
the Stokes parameters transform as
\begin{equation} \label{eqn:rotationStokes}
\arraycolsep=1.4pt\def\arraystretch{1.8}
\begin{array}{ccl}
I_\mathrm{i} &=& \bar{\mathcal{I}_\mathrm{i}} \\
Q_\mathrm{i} &=& \bar{\mathcal{Q}}_\mathrm{i}\cos(2\alpha_\mathrm{i}) +
\bar{\mathcal{U}}_\mathrm{i}\sin(2\alpha_\mathrm{i})\\
U_\mathrm{i} &=& \bar{\mathcal{U}}_\mathrm{i}\cos(2\alpha_\mathrm{i}) -
\bar{\mathcal{Q}}_\mathrm{i}\sin(2\alpha_\mathrm{i})\,.
\end{array}
\end{equation}
Since photons emitted by the star surface, or, more generally,
inside the adiabatic region, are 100\% polarized either in the X or O
mode (i.e. $\bar{\mathcal{Q}_\mathrm{i}}=\pm1$ and $\bar{\mathcal{U}_\mathrm{i}}=0$), 
the Stokes parameters of the radiation collected at infinity are
\begin{equation} \label{eqn:totalStokes}
\arraycolsep=1.4pt\def\arraystretch{2.0}
\begin{array}{ccll}
Q &=& \sum_{\mathrm{i}=1}^N Q_\mathrm{i} &=
\sum_{\mathrm{i}=1}^{N_\mathrm{X}}
\cos(2\alpha_\mathrm{i}) - \sum_{\mathrm{j}=1}^{N_\mathrm{O}}
\cos(2\alpha_\mathrm{j}) \\
U &=& \sum_{\mathrm{i}=1}^N U_\mathrm{i} &=
\sum_{\mathrm{j}=1}^{N_\mathrm{O}}
\sin(2\alpha_\mathrm{j}) - \sum_{\mathrm{i}=1}^{N_\mathrm{X}}
\sin(2\alpha_\mathrm{i})
\end{array}
\end{equation}
where $N_\mathrm{X}$ ($N_\mathrm{O}$) is the number of
extraordinary (ordinary) photons, $N=N_\mathrm{X}+N_\mathrm{O}$,
and we used equations (\ref{eqn:rotationStokes}).

\subsection{Polarization observables}
\label{subsec:polarizationobservables}

The polarization state of the detected radiation can be described
in terms of two observables\footnote{As mentioned earlier,
circular polarization is not considered in the present work.}, the linear
polarization fraction $\Pi_\mathrm{L}$ and the polarization angle
$\chi_\mathrm{p}$ defined as
\begin{equation} \label{eqn:polobs}
\arraycolsep=1.4pt\def\arraystretch{2.2}
\begin{array}{cll}
\Pi_\mathrm{L} &=& \dfrac{\sqrt{Q^2 + U^2}}{I} \\
\chi_\mathrm{p} &=& \dfrac{1}{2}\arctan\left(\dfrac{U}{Q}\right)\,.
\end{array}
\end{equation}
The linear polarization fraction
is not, in general, equivalent to the ratio
$|N_\mathrm{X}-N_\mathrm{O}|/N$ \cite[as previously noticed by][]{Heyletal2003}.
This would happen only if all the angles $\alpha_\mathrm{i}$ were
the same, i.e. when the magnetic field is uniform across the 
emitting region (as in the
case considered by \citealt{LaiHo2003} and \citealt{vanAdelsbergPerna}
of radiation coming from a small hot spot on the NS surface). In
fact, denoting with $\alpha_0$ the common value and substituting
expressions (\ref{eqn:totalStokes}) in the first of equations
(\ref{eqn:polobs}), one obtains
\begin{flalign} \label{eqn:PLunifB}
\Pi_\mathrm{L} &= \dfrac{\sqrt{(N_\mathrm{X}-N_\mathrm{O})^2
\cos^2(2\alpha_0)+(N_\mathrm{O}-N_\mathrm{X})^2\sin^2(2\alpha_0)}}{N} &
\notag \\
\ &= \dfrac{|N_\mathrm{X}-N_\mathrm{O}|}{N}\,. &
\end{flalign}
Under the same hypothesis the polarization angle, given by the
second of equations (\ref{eqn:polobs}), is directly related to the
angle $\alpha_0$
\begin{flalign} \label{eqn:chiunifB}
\chi_\mathrm{p} &= \dfrac{1}{2}\arctan\bigg[\dfrac{(N_\mathrm{O}-
N_\mathrm{X})\sin(2\alpha_0)}{(N_\mathrm{X}-N_\mathrm{O})
\cos(2\alpha_0)}\bigg] = -\alpha_0\,. &
\end{flalign}
Hence,  the polarization fraction gives direct information about the intrinsic degree of
polarization of the radiation (i.e. that at the source) only for a constant rotation angle $\alpha_0$. Under the same conditions, the
polarization angle provides the direction of the
(uniform) magnetic field of the source in the plane of the sky.

On the contrary, if the $B$-field is non-uniform (e.g. for emission 
coming from the entire surface of a NS endowed with a dipole field), 
$\alpha$ will vary according to
the magnetic field direction at the point where the photon crosses
the adiabatic radius. In this case equations (\ref{eqn:totalStokes})
and (\ref{eqn:polobs}) give
\begin{flalign} \label{eqn:PLgen}
\Pi_\mathrm{L} &= \frac{1}{N}\bigg[ N +
2\sum\nolimits_{\mathrm{i}}\sum\nolimits_{\mathrm{k}>\mathrm{i}}
\cos(2\alpha_\mathrm{i}-2\alpha_\mathrm{k}) & \notag \\
\ &\ \ \ \ \ \ \ \ \ \
+ 2\sum\nolimits_\mathrm{j}\sum\nolimits_{\mathrm{r}>\mathrm{j}}
\cos(2\alpha_\mathrm{j}-2\alpha_\mathrm{r}) & \notag \\
\ & \ \ \ \ \ \ \ \ \ \ -2\sum\nolimits_{\mathrm{i}}
\sum\nolimits_{\mathrm{j}}\cos(2\alpha_\mathrm{i}-2\alpha_\mathrm{j})
\bigg]^{1/2}\,,
\end{flalign}
where $\mathrm{i},\mathrm{k}=1,...,N_\mathrm{X}$ and $\mathrm{j},
\mathrm{r}=1,...,N_\mathrm{O}$, while the polarization angle
results in
\begin{flalign} \label{eqn:chigen}
\chi_\mathrm{p} &= \frac{1}{2}\arctan\bigg[
-\frac{\sum_{\mathrm{i}=1}^{N_\mathrm{X}}\sin(2\alpha_\mathrm{i}) -
\sum_{\mathrm{j}=1}^{N_\mathrm{O}}\sin(2\alpha_\mathrm{j})}
{\sum_{\mathrm{i}=1}^{N_\mathrm{X}}\cos(2\alpha_\mathrm{i}) -
\sum_{\mathrm{j}=1}^{N_\mathrm{O}}\cos(2\alpha_\mathrm{j})}\bigg]\,. &
\end{flalign}
So, in the general case both $\Pi_\mathrm{L}$ and $\chi_\mathrm{p}$
depend on the distribution of the angles $\alpha_\mathrm{i}$, which, in
turn, is determined by the geometry of the magnetic field. 

\section{Polarization of surface emission from neutron stars} \label{sec:alphadistribution}
In this section we present quantitative results for the
polarization observables in the case of surface (thermal) emission
from a neutron star endowed with an axially-symmetric magnetic
field, either a dipole or a (globally) twisted dipole, the latter often
used to describe the magnetosphere of magnetars \cite[see][]{TLK}.

\subsection{The $\alpha$--distribution}
\label{subsec:calculations} 
\begin{figure*}
\begin{center}
\includegraphics[width=15cm]{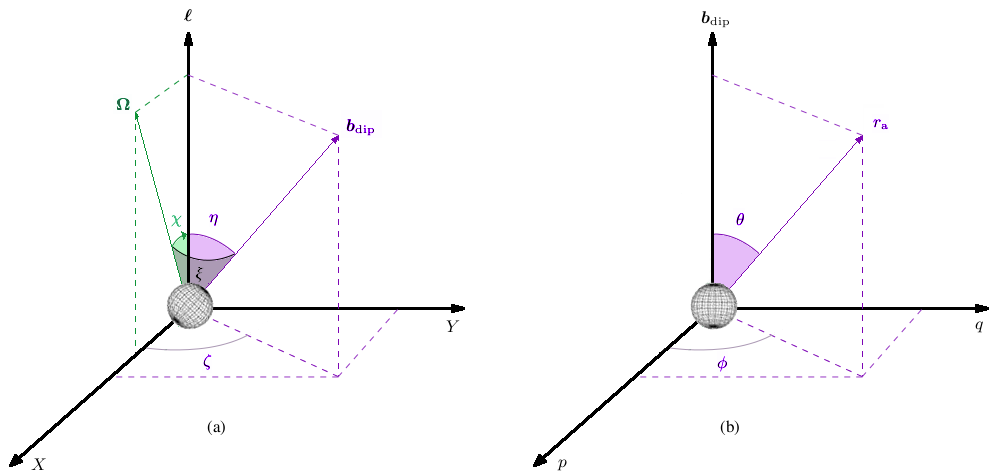}
\caption{The two reference frames used in the calculation of the $\alpha$ angle. {\it Left:} the $(X,Y,Z)$ frame with the
 $Z$ axis in the direction of the LOS $\boldsymbol{\ell}$, the $X$ axis in the
plane $\boldsymbol{\ell}$-$\boldsymbol{\Omega}$, where
$\boldsymbol{\Omega}$ is the star spin axis, and the $Y$ axis
perpendicular to both $X$ and $Z$. {\it Right:} the $(p,q,t)$
reference frame with the $t$ axis
 along the star magnetic axis $\boldsymbol{b}_\mathrm{dip}$, the
$p$ and $q$ two mutually orthogonal axes  in the plane
perpendicular to $\boldsymbol{b}_\mathrm{dip}$ (see Appendix
\ref{subsec:pqtframe} for more details). The angles $\xi$, $\chi$, $\eta$, $\zeta$, $\theta$ and $\phi$ are also shown.}
\label{fig:LOSbdipframe}
\end{center}
\end{figure*}

Let us introduce a reference frame
$(X,Y,Z)$ with the $Z$-axis in the direction of the LOS (unit
vector $\boldsymbol{\ell}$), $\boldsymbol{X}$  in the plane of $\boldsymbol{\ell}$ and the star spin axis (unit vector
$\boldsymbol{\Omega}$) and $\boldsymbol{Y}=\boldsymbol{\ell}\times\boldsymbol{X}$. The geometry is shown in Figure
\ref{fig:LOSbdipframe}a, where $\chi$ is the angle between the
spin axis and the LOS, and $\xi$ is the angle between the spin
axis and the magnetic dipole axis (unit vector
$\boldsymbol{b}_\mathrm{dip}$).
It is $\boldsymbol{\Omega}=(-\sin\chi,0, \cos\chi)$ while, having
introduced the polar angles $\eta$ and $\zeta$ that fix
the direction of $\boldsymbol{b}_\mathrm{dip}$ with respect to
$\boldsymbol{\ell}$ (see again Figure \ref{fig:LOSbdipframe}a), one
has $\boldsymbol{b}_\mathrm{dip}=
(\sin\eta\cos\zeta,\sin\eta\sin\zeta,\cos\eta)$. The angles $\eta$
and $\zeta$ are related to $\chi$ and $\xi$ by
\begin{equation} \label{eqn:etazeta}
\arraycolsep=1.4pt\def\arraystretch{1.8}
\begin{array}{ccl}
\cos\eta &=& \cos\chi\cos\xi + \sin\chi\sin\xi\cos\gamma \\
\cos\zeta &=& \dfrac{\cos\xi - \cos\chi\cos\eta}{\sin\chi\sin\eta}\,,
\end{array}
\end{equation}
where $\gamma$ is the rotational phase. Using the previous
expressions, the components of the unit vector
$\boldsymbol{b}_\mathrm{dip}$ in the frame $(X,Y,Z)$ become
\begin{equation} \label{eqn:bdip}
\boldsymbol{b}_\mathrm{dip} = \left(\begin{array}{c}
\sin\chi\cos\xi - \cos\chi\sin\xi\cos\gamma \\
\sin\xi\sin\gamma \\
\cos\chi\cos\xi + \sin\chi\sin\xi\cos\gamma
\end{array}\right)\,.
\end{equation}

According to the discussion in \S \ref{subsec:stokesparameters},
the axes $u$ and $v$ of the polarimeter frame can be
chosen as any pair of orthogonal directions in the $XY$ plane.
In general it is
\begin{align} \label{eqn:baseu}
\boldsymbol{u} \equiv \boldsymbol{X} = \left(\begin{array}{c}
\cos\psi \\
\sin\psi \\
0
\end{array}\right) &\,,& \boldsymbol{v} \equiv \boldsymbol{Y} =
\left(\begin{array}{c}
-\sin\psi \\
\cos\psi \\
0
\end{array}\right)\,.
\end{align}
where $\psi$ is the angle that the $u$ axis makes with the
$X$ axis. The axes $x$ and $y$ of the reference frame $(x,y,z)$, that change
for each photon, are defined once the magnetic field geometry is
fixed as
\begin{align} \label{eqn:basexdef}
\boldsymbol{x} = \frac{\boldsymbol{\ell}\times\boldsymbol{B}}
{|\boldsymbol{\ell}\times\boldsymbol{B}|} &\,,& \boldsymbol{y} =
\boldsymbol{\ell}\times\boldsymbol{x}\,.
\end{align}
The angle $\alpha$ by which the photon frame $(x,y,z)$ has to be
rotated to coincide with the polarimeter frame $(u,v,w)$ is then
simply obtained taking the scalar product of $\boldsymbol{u}$ with
$\boldsymbol{x}$
\begin{equation} \label{eqn:alphadef}
\cos\alpha = \boldsymbol{u}\cdot\boldsymbol{x}\,.
\end{equation}
The indetermination in the sign of $\alpha$ is resolved looking at
the sign of $\boldsymbol{v}\cdot\boldsymbol{x}$; if the latter is
positive the rotation is by an angle $-\alpha$ (i.e.
$\sin\alpha=-\sqrt{1-\cos^2\alpha}$).

Since we need to  consider photons only from the boundary of the
adiabatic region outwards, $\boldsymbol{B}$ in the first of
equations (\ref{eqn:basexdef}) is the stellar magnetic field
calculated at $r_\mathrm{a}$ (see equation \ref{eqn:ra}).
Actually, it is more convenient to express the magnetic field
components in a reference frame $(p,q,t)$, with the $t$ axis along
$\boldsymbol{b}_\mathrm{dip}$ and $p,q$  two mutually orthogonal
directions in the plane perpendicular to $t$ (see Figure
\ref{fig:LOSbdipframe}b). For a dipole, in particular, the
polar components of the magnetic field in this frame are
\begin{flalign} \label{eqn:Bdippolar}
\boldsymbol{B}^\mathrm{pol} &= \left(\begin{array}{c}
B_r \\
B_\theta \\
B_\phi
\end{array}\right) = \frac{B_\mathrm{P}}{2}\left(\frac{R_\mathrm{NS}}
{r_\mathrm{a}}\right)^3\left(\begin{array}{c}
2\cos\theta \\
\sin{\theta} \\
0
\end{array}\right)\,, &
\end{flalign}
where $\theta$ is the magnetic colatitude. Then, the cartesian
components $\boldsymbol{B}=(B_p,B_q, B_t)$ can be calculated
making use of expressions (\ref{eqn:Bdipcartes}) in Appendix
\ref{subsec:Bfield}. However, since all the calculations to derive
the analytical expression of the angle $\alpha$ (see equation
\ref{eqn:alphadef}) are in the LOS reference frame, the $X$, $Y$
and $Z$ components of $\boldsymbol{B}$ are needed. They can be
obtained through a change of basis as
\begin{flalign} \label{eqn:changeofbase}
B_X &= B_pp_X + B_qq_X + B_t(b_\mathrm{dip})_X & \notag \\
B_Y &= B_pp_Y + B_qq_Y + B_t(b_\mathrm{dip})_Y & \\
B_Z &= B_pp_Z + B_qq_Z + B_t(b_\mathrm{dip})_Z\,, & \notag
\end{flalign}
where the components of $\boldsymbol{p}$, and $\boldsymbol{q}$ in
the $(X,Y, Z)$ frame are given in Appendix \ref{subsec:pqtframe},
while those of $\boldsymbol{b}_{dip}$ are given by equation
(\ref{eqn:bdip}).

\begin{figure}
\begin{center}
\includegraphics[width=7cm]{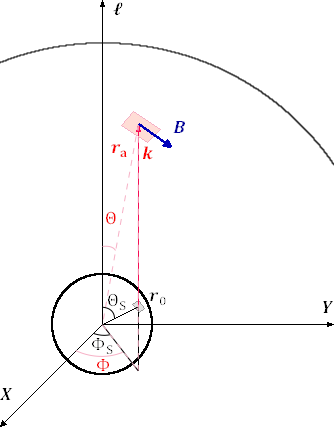}
\caption{The path of a photon emitted by a point on the star surface
with  polar angles $\Theta_\mathrm{S}$ and
$\Phi_\mathrm{S}$, that crosses the adiabatic boundary in a point
of polar coordinates  $\Theta$ and $\Phi$.}
\label{fig:adiabregion}
\end{center}
\end{figure}

Substituting the expressions (\ref{eqn:changeofbase}) in the first
of equations (\ref{eqn:basexdef}), the components of the unit
vector $\boldsymbol{x}$ in the LOS reference frame are
\begin{flalign} \label{eqn:basex}
\boldsymbol{x} &= \frac{1}{\sqrt{B_X^2+B_Y^2}}\left(\begin{array}{c}
-B_Y \\
B_X \\
0
\end{array}\right)\,. &
\end{flalign}
$B_X$ and $B_Y$ clearly depend on the angles $\chi$, $\xi$ and the phase
$\gamma$ through the unit vectors of the $(p,q,t)$ reference
frame, given by equations (\ref{eqn:pXYZ}), (\ref{eqn:qXYZ}) and
(\ref{eqn:bdip}). Moreover, they depend also on the magnetic
colatitude and azimuth ($\theta$ and $\phi$ that fix the point
where the magnetic field is calculated on the adiabatic surface)
through the components $B_p$, $B_q$ and $B_t$ given by equations
(\ref{eqn:Bdipcartes}) and (\ref{eqn:Bdippolar}). Actually, the
angles $\theta$ and $\phi$ depend in turn on $\chi$, $\xi$ and
$\gamma$. To make  this dependence explicit, let us consider
Figure \ref{fig:adiabregion}, that shows, in the LOS reference
frame, the path of a photon emitted from a point of the surface
characterized by the polar angles $\Theta_\mathrm{S}$ and
$\Phi_\mathrm{S}$, up to the point where it crosses the adiabatic
boundary, characterized by the angles $\Theta$ and $\Phi$.
Observing the star at infinity, one collects only photons that
travel along vectors $\boldsymbol{k}=(0,0,k)$ parallel to the LOS
$\boldsymbol{\ell}$. The modulus of each vector $\boldsymbol{k}$
is fixed by the condition:
\begin{equation} \label{eqn:conditionk}
\boldsymbol{r}_0 + \boldsymbol{k} = \boldsymbol{r}_\mathrm{a}\,,
\end{equation}
where $\boldsymbol{r}_0=R_\mathrm{NS}(\sin\Theta_\mathrm{S}
\cos\Phi_\mathrm{S},\sin\Theta_\mathrm{S}\sin\Phi_\mathrm{S},
\cos\Theta_\mathrm{S})$ is the position vector of the surface
point from which the photon has been emitted and
$\boldsymbol{r}_\mathrm{a}=
r_\mathrm{a}(\sin\Theta\cos\Phi,\sin\Theta\sin\Phi,\cos\Theta)$ is
the position vector of the point where the photon crosses
the adiabatic boundary. Taking the norm of both the sides of
equation (\ref{eqn:conditionk}) and solving for $k$, the only
acceptable solution is
\begin{equation} \label{eqn:modulek}
k = \sqrt{r_\mathrm{a}^2-R_\mathrm{NS}^2\sin^2\Theta_\mathrm{S}} -
R_\mathrm{NS}\cos\Theta_\mathrm{S}
\end{equation}
and, substituting this result again in equation (\ref{eqn:conditionk}),
one obtains:
\begin{equation} \label{eqn:ravector}
\boldsymbol{r}_\mathrm{a} = \left(\begin{array}{c}
R_\mathrm{NS}\sin\Theta_\mathrm{S}\cos\Phi_\mathrm{S} \\
R_\mathrm{NS}\sin\Theta_\mathrm{S}\sin\Phi_\mathrm{S} \\
\sqrt{r_\mathrm{a}^2-R_\mathrm{NS}^2\sin^2\Theta_\mathrm{S}}
\end{array}\right)\,,
\end{equation}
where the distance $r_\mathrm{a}$ of the adiabatic boundary is
given by equation (\ref{eqn:ra}). From simple geometrical
considerations (see again Figure \ref{fig:LOSbdipframe}b), it
follows that
\begin{equation} \label{eqn:costheta}
\cos\theta = \boldsymbol{b}_\mathrm{dip}\cdot
\frac{\boldsymbol{r}_\mathrm{a}}{r_\mathrm{a}}\,,
\end{equation}
while the cosine of the angle $\phi$ can be obtained as
\begin{equation} \label{eqn:cosphi}
\cos\phi = \boldsymbol{p}\cdot\boldsymbol{r}_\mathrm{a}^\perp\,,
\end{equation}
where $\boldsymbol{r}_{\mathrm{a}}^\perp$ is the unit vector of the
projection of $\boldsymbol{r}_\mathrm{a}$ orthogonal to
$\boldsymbol{b}_\mathrm{dip}$. The complete expressions of $\cos\theta$
and $\cos\phi$ are given in Appendix \ref{subsec:costhetacosphi}.

Finally, substituting into equation (\ref{eqn:alphadef}) gives the
distribution of $\alpha$
\begin{equation} \label{eqn:cosalpha}
\cos\alpha =
\frac{B_X\sin\psi - B_Y\cos\psi}{\sqrt{B_X^2+B_Y^2}}\,,
\end{equation}
which is a function of the angles $\chi$, $\xi$, the phase $\gamma$,
the photon energy $E$ and $B_\mathrm{P}$ (through the adiabatic radius
$r_\mathrm{a}$, see equation \ref{eqn:ra}), the polar angles
$\Theta_\mathrm{S}$ and $\Phi_\mathrm{S}$ that fix the point on
the surface from which the photons were emitted and the angle $\psi$ by 
which the polarimeter frame is rotated wrt the LOS one. In the following we take
$\psi=0$, i.e. the $u$ ($v$) axis coincides with the $X$ ($Y$) axis, 
although the generalization to other values is straightforward.

\subsection{Numerical implementation}
\label{subsec:numericalimplementation}
\begin{figure*}
\begin{center}
\includegraphics[width=17.5cm]{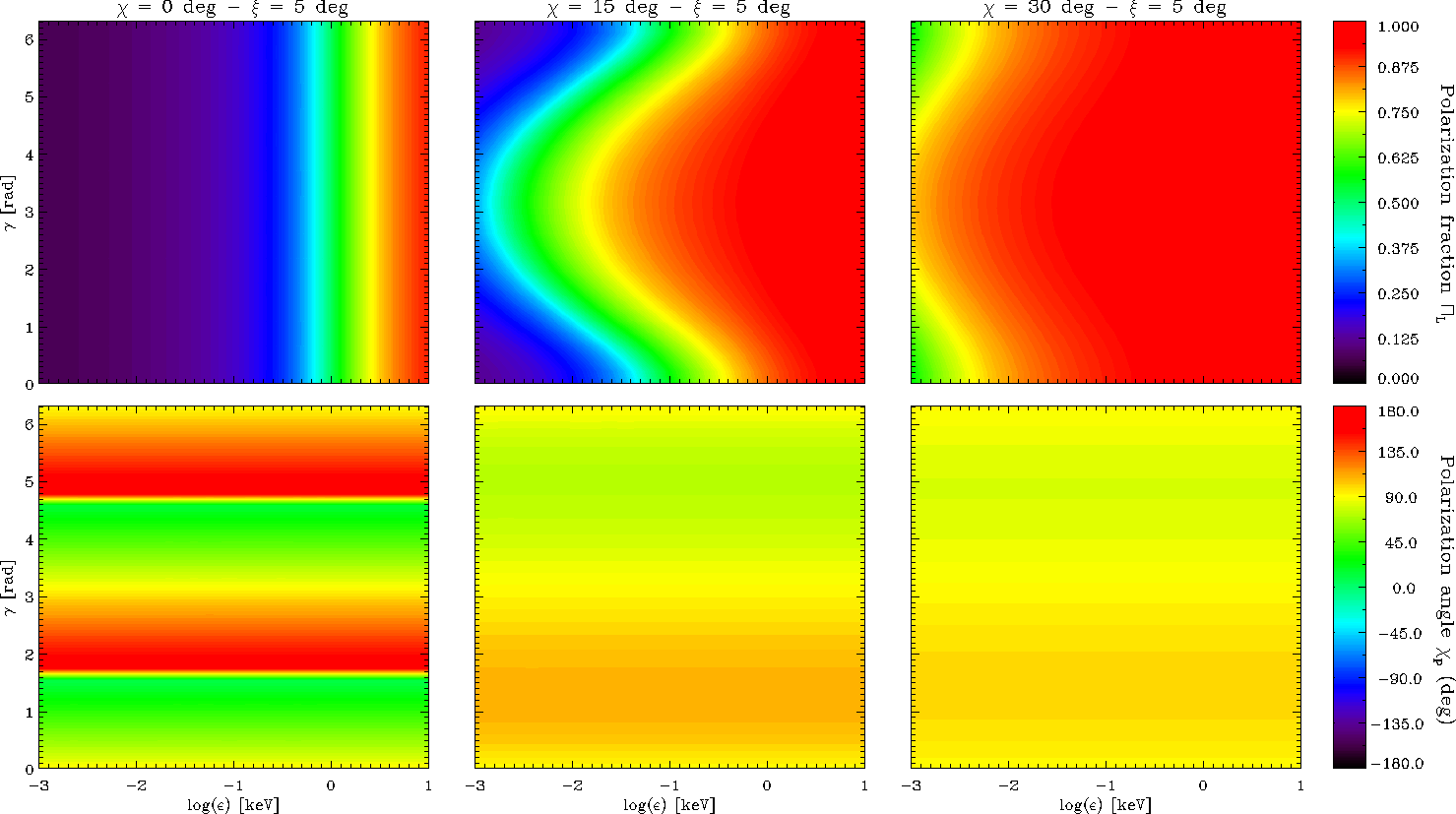}
\caption{Contour plots in the energy-phase plane of the polarization fraction (top row) and the
polarization angle (bottom row) for a neutron star with
$B\mathrm{P}=10^{13}$ G, $R_\mathrm{NS}=10$ km and mass $M_\mathrm{NS}
=1.4\,M_\odot$. The inclination of the magnetic axis with respect
to the spin axis is fixed to $\xi=5^\circ$, while the angle between the spin axis and the LOS is
$\chi=0^\circ$ (left column),
$15^\circ$ (middle column) and $30^\circ$ (right column). Seed photons
are 100\% polarized in the X-mode.}
\label{fig:plot1}
\end{center}
\end{figure*}

In order to calculate the polarization fraction
$\Pi_\mathrm{L}$ and the polarization angle $\chi_\mathrm{p}$, we use
the ray-tracing code developed by \citet{ZaneTurolla}, with the addition
of a specific module for the evaluation of the $\alpha$ angle
distribution and of the Stokes parameters. QED effects are included as described in section
\ref{sec:theoreticaloverview}.
The code takes also into account the effects due to the
strong gravity on photon propagation (relativistic ray-bending) and on the 
stellar magnetic field. For a dipole field (see equations \ref{eqn:Bdippolar}), 
the latter are given by
\begin{flalign} \label{eqn:BdippolarGR}
\arraycolsep=1.4pt\def\arraystretch{1.8}
\begin{array}{l}
B_r^\mathrm{GR} = f_\mathrm{dip}B_r \\
B_\theta^\mathrm{GR} = g_\mathrm{dip}B_\theta \\
B_\phi^\mathrm{GR} = B_\phi = 0\,,
\end{array} & &
\end{flalign}
where
\begin{flalign} \label{eqn:fdipgdip}
\arraycolsep=1.4pt\def\arraystretch{1.8}
\begin{array}{l}
f_\mathrm{dip} =-\dfrac{3}{x^3}\left[\ln(1-x)+\dfrac{1}{2}x(x+2)
\right] \\
g_\mathrm{dip} =\sqrt{1-x}\left(-2f_\mathrm{dip}+\dfrac{3}{1-x}\right)\,,
\end{array} & &
\end{flalign}
with $x=R_\mathrm{s}/r$; $R_\mathrm{s}=2GM_\mathrm{NS}/c^2$ is the
Schwarzschild radius and $M_\mathrm{NS}$ is the stellar mass \cite[see][]{PageSarmiento}.

The expressions for the total Stokes parameters $Q$
and $U$ given in equations (\ref{eqn:totalStokes}) can be easily generalized to a continuous photon distribution
by replacing the sums with integrals over the visible part of the star surface
\begin{equation}\label{eqn:integralStokes}
\arraycolsep=1.4pt\def\arraystretch{2.2}
\begin{array}{ccl}
F_Q &=& \displaystyle\int_0^{2\pi}\mathrm{d}\Phi_\mathrm{S}
\int_0^1 \mathrm{d}u^2 \big(n_\mathrm{X}-n_\mathrm{O}\big)
\cos(2\alpha) \\
F_U &=& \displaystyle\int_0^{2\pi}\mathrm{d}\Phi_\mathrm{S}
\int_0^1 \mathrm{d}u^2 \big(n_\mathrm{O}-n_\mathrm{X}\big)
\sin(2\alpha)\, ,
\end{array}
\end{equation}
where $n_\mathrm{X}$ ($n_\mathrm{O}$) is the photon intensity 
in the extraordinary (ordinary) mode and $F_Q$ and $F_U$ are the 
``fluxes'' of the Stokes parameters \cite[see][]{PavlovZavlin}.
In general, $n_\mathrm{X}$ and $n_\mathrm{O}$ depend on the photon
energy $E$ and direction, and on the position on the star surface of the
emission point.
The integration variable $u=\sin\bar\Theta$ is related to
$\Theta_\mathrm{S}$ by the integral \cite[see][and references
therein]{ZaneTurolla}:
\begin{flalign} \label{eqn:raybending}
\bar\Theta &= \int_0^{1/2}\frac{\mathrm{d}v\sin\Theta_\mathrm S}{\left[
\left(1-{x}\right)/4-\left(1-{2v}{x}\right)v^2\sin^2\Theta_\mathrm S
\right]^{1/2}}\,, &
\end{flalign}
that accounts for ray-bending and reduces to $\bar\Theta=\Theta_\mathrm{S}$
in the limit $x\rightarrow 0$ (when the effects of general relativity
can be neglected). The total photon flux is
obtained in a similar way
\begin{equation} \label{eqn:integralI}
\begin{array}{ccl}
F_I &=& \displaystyle\int_0^{2\pi}\mathrm{d}\Phi_\mathrm{S}
\int_0^1 \mathrm{d}u^2 \big(n_\mathrm{X}+n_\mathrm{O}\big)\,.
\end{array}
\end{equation}

For the sake of simplicity, we assume in the following that
radiation is emitted by the cooling star surface with an isotropic
blackbody distribution. The photon intensity is then
\begin{equation} \label{eqn:IBB}
n_\mathrm{X,O} = \frac{2}{h^2c^2}
\frac{E^2}{\mathrm{e}^{E/kT}-1}\,,
\end{equation}
where $T$ is the local surface temperature. In order to model
the surface thermal distribution and to avoid a vanishing temperature
at the equator, here we adopt a variant
of the standard temperature distribution for a
core-centred dipole field \cite[e.g.][]{Page95},
$T(\vartheta)=\max(T_\mathrm p\vert\cos\vartheta\vert^{1/2},T_\mathrm
e)$, where $\vartheta$ is the angle between the local normal and $\mathbf B$, $T_\mathrm p$ and
$T_\mathrm e$ are the temperature at the pole and at the equator,
respectively; in the following we take $T_\mathrm p=150$ eV and $T_\mathrm e=100$ eV. The polarization degree of the radiation emitted at
the surface is fixed specifying the ratio $p_0=n_\mathrm{X}/(n_\mathrm{X}+n_\mathrm{O})$,
$\vert n_\mathrm{X}-n_\mathrm{O}\vert/(n_\mathrm{X}+n_\mathrm{O})=\vert 2p_0-1\vert$.

\subsection{Results}
\label{subsec:simulationsresults}
\begin{figure}
\begin{center}
\includegraphics[width=8.5cm]{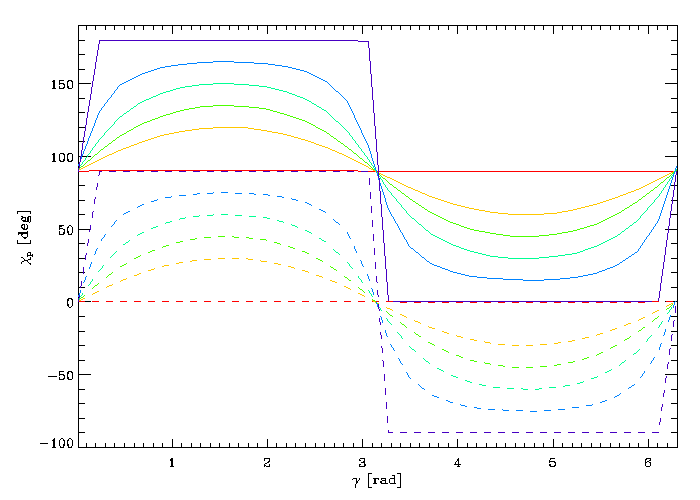}
\caption{Polarization angle as a function of the rotational
phase at a fixed energy ($E=0.02$ keV), for $\chi=90^\circ$ and different
values of $\xi$: $0^\circ$ (red), $30^\circ$ (orange), $45^\circ$ (green),
$60^\circ$ (light blue),  $75^\circ$ (blue) and $90^\circ$ (violet). The solid (dashed) lines
correspond to seed photons 100\% polarized in the X-mode (O-mode). The
values of $R_\mathrm{NS}$, $M_\mathrm{NS}$ and $B_\mathrm{P}$ are the
same as in Figure \ref{fig:plot1}.}
\label{fig:chipolXO}
\end{center}
\end{figure}
The polarization observables $\Pi_\mathrm{L}$ and
$\chi_\mathrm{p}$ can be computed recalling the definitions given
in equation (\ref{eqn:polobs}) and using the expressions we just
derived for the Stokes parameters, equations
(\ref{eqn:integralStokes}) and (\ref{eqn:integralI}), together
with the distribution of $\cos\alpha$ given in equation
(\ref{eqn:cosalpha}). All results presented in this section refer
to a neutron star with mass $M_{\mathrm {NS}}=1.4M_\odot$ and
radius $R_{\mathrm {NS}}=10\, {\rm km}$. Thermal photons are
assumed to be $100$\% polarized in one of the two modes, i.e. 
$p_0=0,1$; essentially we consider all the photons as
extraordinary, unless explicitly stated otherwise \cite[see e.g. the discussion
in][]{Fernandez+Davis,Tavernaetal}. This is the choice 
which produces the most unfavourable conditions to detect the 
depolarizing effects of vacuum polarization and geometry on the 
polarization observables.

Figure \ref{fig:plot1} shows the polarization fraction and the
polarization angle as functions of the photon energy and the
rotational phase for different values of the inclination $\chi$ of
the LOS wrt the star spin axis. The magnetic axis is at an angle
$\xi=5^\circ$ with respect to the spin axis (i.e. the NS is a
nearly aligned rotator) and $B_\mathrm{P}=10^{13}$ G. The effects
produced by the frame rotation (induced by the non-uniform $B$-field)
are quite dramatic, as it is evident from the polarization
fraction (top row). In particular, for $\chi=0$ (top left panel)
$\Pi_\mathrm{L}$ is almost everywhere far from unity, the value expected
from the intrinsic degree of polarization, $\vert
n_\mathrm{X}-n_\mathrm{O}\vert/(n_\mathrm{X}+n_\mathrm{O})=1$, and it
becomes $\sim 0.9$ only at $E\sim 10$ keV. By
increasing the LOS inclination ($\chi=15^\circ$, top middle
panel), the polarization fraction reaches unity for photon
energies $\ga 1\ {\mathrm {keV}}$, while at lower energies it is
substantially smaller (between $\sim 0.1$ and $\sim 0.8$).
Only when $\chi$ becomes sufficiently large ($\chi=30^\circ$, top
right panel) $\Pi_\mathrm{L}$ is unity, except at
low energies ($\sim 1$--10 eV), where the polarization
fraction drops to about 0.6 in some phase intervals.

The bottom row of Figure \ref{fig:plot1} shows $\chi_\mathrm{p}$
for the same three simulations. Contrary of what happens for the
polarization fraction, the polarization angle does not depend on
the energy and exhibits an oscillatory behavior as a function of
the rotational phase around a value of $90^\circ$. The amplitude
of the oscillations depends on the geometrical angles and, for some
combinations of $\chi$ and $\xi$, $\chi_\mathrm{p}$ sweeps the
entire range $[0^\circ,\,180^\circ]$ through a discontinuity, or ``jump''. This is
clearly seen in the bottom left panel of Figure \ref{fig:plot1}
where $\chi=0$ , while $\chi_\mathrm{p}$ is in between $\sim
70^\circ$--$110^\circ$ (bottom middle panel) and $\sim
80^\circ$--$100^\circ$ (bottom right panel) for $\chi=15^\circ$
and $\chi=30^\circ$, respectively. This is further illustrated in
Figure \ref{fig:chipolXO}, which shows the polarization angle as a
function of the rotational phase at a single energy ($E=0.02\
\mathrm{keV}$), $\chi=90^\circ$ and different values of $\xi$ for
radiation 100\% polarized in the X-mode (solid lines) and in the
O-mode (dashed lines). The amplitude of the oscillation vanishes
in the case of an aligned rotator seen equator-on
($\chi=90^\circ$, $\xi=0^\circ$) and increases for increasing
$\xi$ until the ``jump'' appears for $\xi=90^\circ$ \footnote{The
curves for $\xi=90^\circ$ in Figure \ref{fig:chipolXO} are
box-like; the sloping lines are an artifact introduced by the
finite resolution of the phase grid.}. The average value of
$\chi_\mathrm{p}$, instead, does not change with $\chi$ and $\xi$
and is fixed by the polarization mode of the seed photons: it is
$90^\circ$ for X-mode photons and $0^\circ$ for O-mode ones
\footnote{Actually the mean value is the same even if photons are
not all polarized in the same mode; since the Stokes parameters
for O- and X-mode photons have opposite signs and the polarization
observables are obtained by summing the Stokes parameters over all
photons, the mean value of $\chi_\mathrm{p}$ reflects the
polarization mode which dominates.}. It should be noted, however,
that the mean value of the polarization angle is not univocally
associated to the two photon modes, since it depends on the choice
of the fixed reference frame $(u,v,w)$, i.e. on the angle
$\psi$ introduced in \S \ref{subsec:calculations}. If, for
instance, $\psi=90^\circ$ (so that the $u$ axis coincides
with the $Y$ axis of the LOS reference frame), the situation depicted in Figure
\ref{fig:chipolXO} is reversed, with the polarization angle for
X-mode photons oscillating around $0^\circ$ and that for the
O-mode ones around $90^\circ$. Of course, different choices of the
$\psi$ angle do not affect the polarization degree $\Pi_\mathrm L$,
the amplitude of the oscillations of $\chi_\mathrm P$ and the
shift of $90^\circ$ between the mean values of $\chi_\mathrm P$
for X-mode and O-mode photons.

\begin{figure*}
\begin{center}
\includegraphics[width=17.5cm]{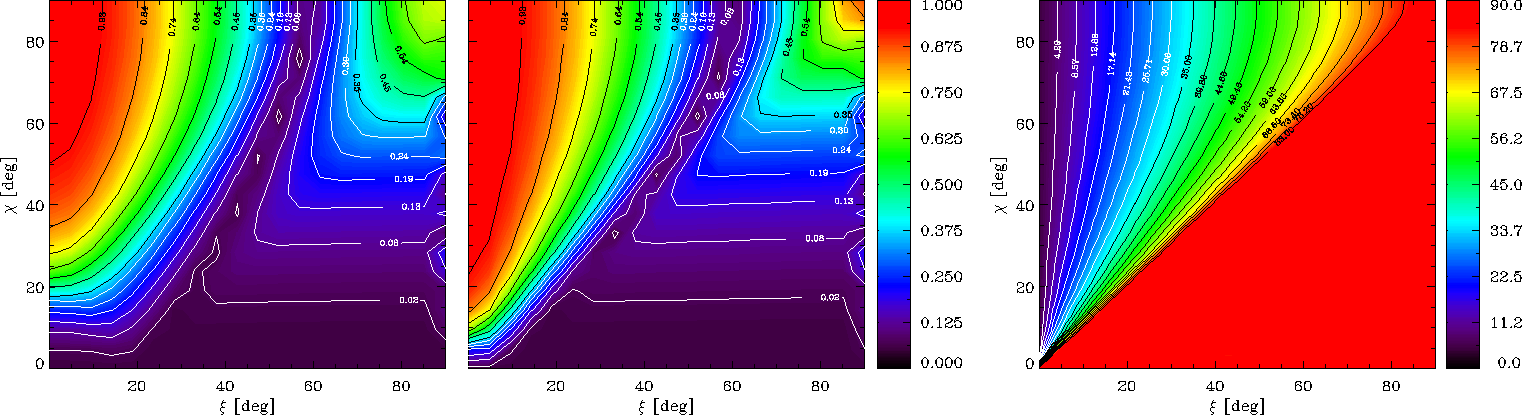}
\caption{Contour plots for the phase-averaged polarization
fraction at optical (2 eV, left panel) and X-ray (0.3 keV, middle panel)
energies, and of the semi-amplitude of the oscillations of the
polarization angle (right panel),  as functions of $\chi$ and
$\xi$. The values of $R_\mathrm{NS}$, $M_\mathrm{NS}$ and
$B_\mathrm{P}$ are the same as in Figure \ref{fig:plot1}.}
\label{fig:plotchicsi}
\end{center}
\end{figure*}

The behaviour of the phase-averaged
polarization fraction as a function of the angles $\chi$, $\xi$ is
shown in Figure \ref{fig:plotchicsi} for two values of the energy,
$E=2$ eV (optical) and $E=0.3$ keV (X-rays). The right panel
illustrates the variation of the semi-amplitude of $\chi_\mathrm
P$. As already noted by \cite{Fernandez+Davis}, the amplitude is
$180^\circ$ for $\xi\la\chi$ when the phase-averaged polarization
degree attains its minimum value (see \S\ref{sec:conclusion}).

\begin{figure*}
\begin{center}
\includegraphics[width=17.5cm]{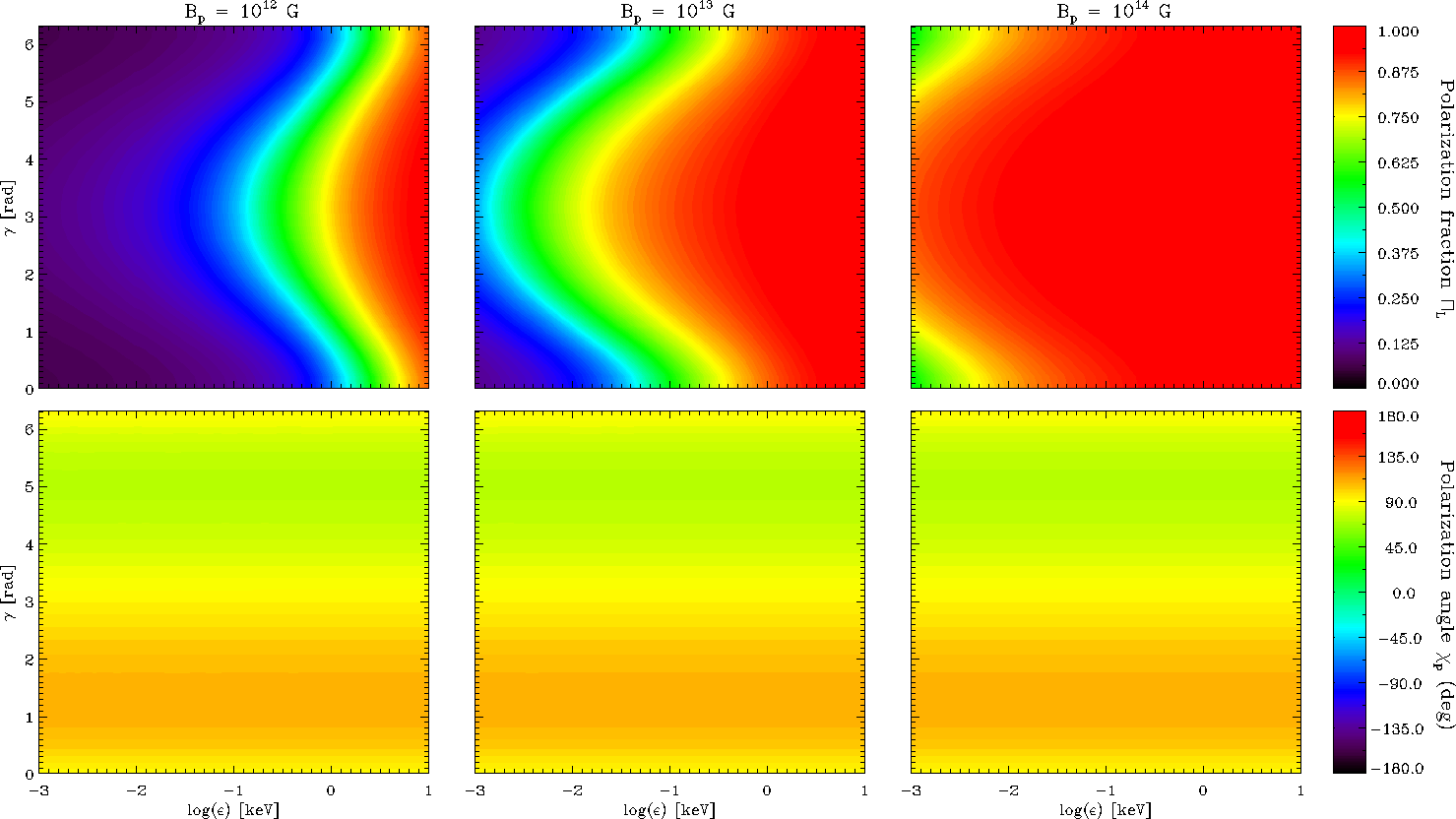}
\caption{Same as in Figure \ref{fig:plot1} for $\chi=15^\circ$, $\xi=5^\circ$ and three different values of the magnetic field:
$B_\mathrm P= 10^{12}$ (left column), $10^{13}$ (middle column) and $10^{14}\, \rm G$ (right colmun).}
\label{fig:plot2}
\end{center}
\end{figure*}

The effects of varying the magnetic field strength are illustrated
in Figure \ref{fig:plot2}, where $\chi=15^\circ$, $\xi=5^\circ$ and
$B_\mathrm{P}=10^{12}$ G (left panel), $B_\mathrm{P}=10^{13}$ G
(middle panel; this is the same case shown in Figure
\ref{fig:plot1}) and $B_\mathrm{P}=10^{14}$ G (right panel).
Again, changes are mostly in the polarization fraction
$\Pi_\mathrm{L}$ (top row). Overall, the polarization fraction is
smaller when the magnetic field is lower (top left panel), and increases for increasing $B_\mathrm
P$, reaching values $\sim 1$ (i.e. the intrinsic polarization degree)
in almost the entire energy range for $B_\mathrm P=10^{14}\ \mathrm
G$ (top right panel), see \S \ref{sec:conclusion}. On the contrary, the polarization angle
(bottom row) does not change much, exhibiting an oscillation
between $\sim 70^\circ$ and $\sim 110^\circ$ at all the values of
$B_\mathrm{P}$.

\begin{figure*}
\begin{center}
\includegraphics[width=13.2cm]{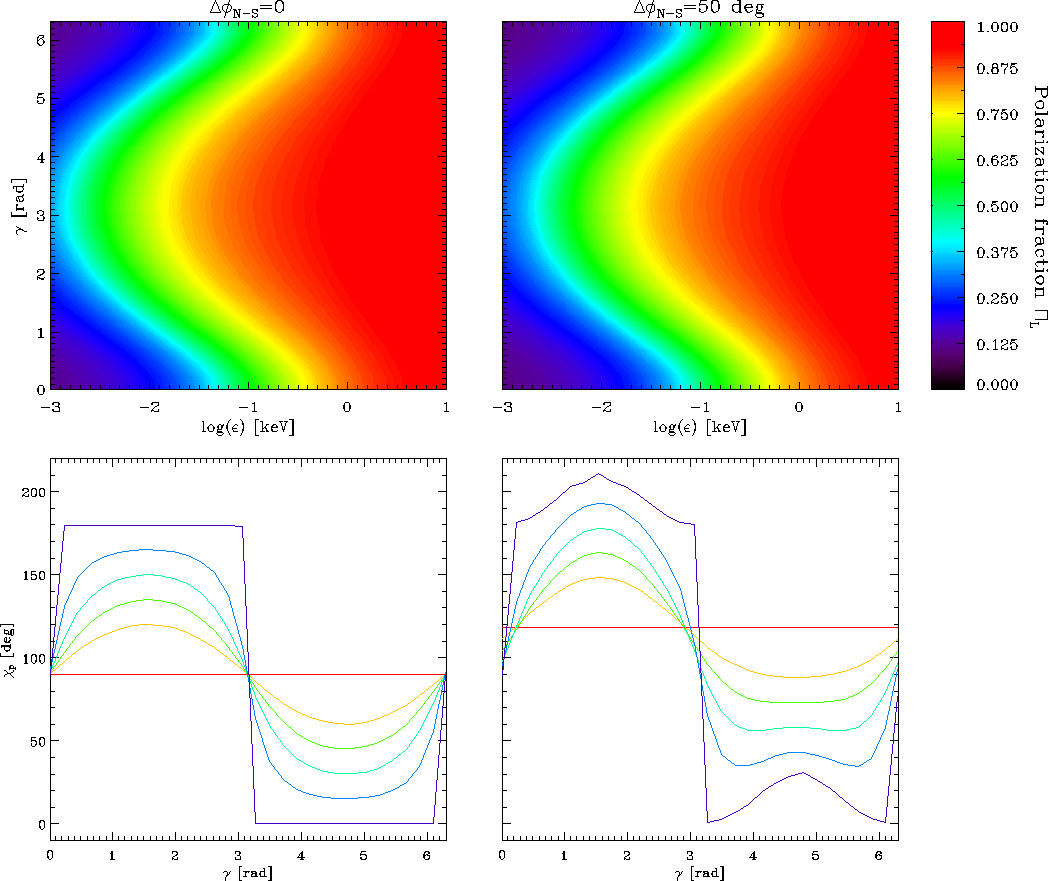}
\caption{Polarization observables for the cases of a pure dipolar
magnetic field (left column) and a globally twisted dipole with twist
angle $\Delta\phi_\mathrm{N-S}\simeq 50^\circ$ (right column). Top row:
polarization fraction in the energy-phase plane for $\chi=15^\circ$ and
$\xi=5^\circ$. Bottom row: polarization angle as a function of the
rotational phase for a fixed photon energy ($E=0.02$ keV),
$\chi=90^\circ$ and $\xi=0^\circ$ (red), $30^\circ$ (orange), $45^\circ$
(green), $60^\circ$ (light blue), $75^\circ$ (blue) and $90^\circ$ (violet). All the plots
are obtained for seed photons 100\% polarized in the X-mode and values
of $R_\mathrm{NS}$, $M_\mathrm{NS}$ and $B_\mathrm{P}$ as in Figure
\ref{fig:plot1}.}
\label{fig:plot4}
\end{center}
\end{figure*}

Finally, Figure \ref{fig:plot4} illustrates the effects on the
polarization observables induced by the presence of a toroidal
field component. The right column shows the phase-energy contour
plot of the polarization fraction (top panel) and a phase plot of
the polarization angle at a fixed energy\footnote{The twisted field actually introduces a dependence of $\chi_\mathrm p$ on the
photon energy. For the values of the twist angle we consider, however, this dependence is quite small.} (bottom panel) for a
globally twisted dipole field. The left column shows for
comparison the same quantities for a pure dipole with the same
$B_\mathrm P=10^{13}\ \mathrm G$. The twisted magnetic field was
evaluated using the analytical approximation by \citet[][see
expressions in their Appendix A]{Pavan}, with a twist angle
$\Delta\phi_\mathrm{N-S}\simeq 50^\circ$. Since relativistic
corrections are unavailable for a twisted field, they were not
applied also to the dipole we show for comparison, whereas
ray bending is still considered in both the cases. The effects of
the twist on the polarization fraction are quite modest and the
variation of $\Pi_\mathrm{L}$ with photon energy and rotational
phase is nearly the same as in the pure dipole case. The only
difference is in a slight overall decrease in the polarization
degree.

The twist of the external field affects much more the polarization
angle, as it can be seen from the bottom row of Figure
\ref{fig:plot4}. The net effect is an overall asymmetry of the oscillations:
$\chi_\mathrm{p}$ sweeps a larger angle in a half-period with respect to the
purely dipolar case; this effect increases with the twist angle
$\Delta\phi_\mathrm{N-S}$, as already noticed by \citet{Fernandez+Davis}.

\section{Discussion and conclusions}
\label{sec:conclusion}

In this paper we reconsidered the problem of the relation between
the intrinsic and observed polarization properties in the case of surface
emission from a neutron star. Our work extends previous investigations 
\cite[]{Heyletal2003,LaiHo2003,vanAdelsbergPerna} by providing the polarization
observables for a large set of physical and geometrical parameters (i.e. 
the angles $\chi$ and $\xi$, the photon energy and the magnetic field strength and topology).
Our treatment includes both ``geometrical'' effects, due to the
rotation of the Stokes parameters which is needed when the magnetic field 
is not constant across the emitting region, and ``vacuum polarization'' 
\cite[][]{HeylShaviv2002};
in order to make a full exploration of the parameter space possible, an
approximated treatment of QED was used. This resulted in much shorter computational
times, without loosing significant physical accuracy, as the comparison with
available results shows \cite[][]{Heyletal2003}. We checked that a typical run 
required about 100 minutes integrating equations (\ref{eqn:odevacuumpol}) and
only few tens of seconds using our approximation. Moreover, our approach
allows to better disentangle the effects of QED and those due to 
the rotation of the Stokes parameters on the polarization signals.
We stress that our main goal was not to compute polarization observables for a precise, 
physical model of surface emission, but to systematically illustrate 
the role of these two effects in making polarization patterns different from
those of the original radiation.
To this end we assumed a quite simple
picture in which the star magnetic field is a dipole and each
surface patch emits a (isotropic) blackbody spectrum at the local
temperature. Our results can be easily generalized to other
magnetic configurations (the case of a globally twisted field is
actually discussed here) and to different surface emission models.
A (comparative) analysis of the polarization observables for emission from an
atmosphere \cite[][and references therein]{vanAdelsbergLai} or from a
condensed surface \cite[][and references therein]{Potekhinetal} will be the
subject of a future paper (Gonzalez Caniulef et al., in preparation).

A crucial point in assessing the measured polarization properties
is that the polarization state of a photon propagating in a
magnetized vacuum (either ordinary or extraordinary) is strictly
related to the choice of the reference frame. In fact, the photon
polarization mode is defined only with respect to the plane fixed by
the wavevector and the local magnetic field. This means that the Stokes parameters of
each photon are in general referred to different
frames, with the two axes orthogonal to the direction of
propagation tailored on the direction of the local $B$-field (\S
\ref{subsec:stokesparameters}). However, photons are collected in
the focal plane of an instrument, where a reference direction has
been a priori introduced.
This means that in order to obtain the
polarization observables relative to the photons received by the
instrument in a given exposure time, the Stokes parameters of each
photon must be transformed to the polarimeter reference frame,
through a rotation in the plane orthogonal to the line-of-sight by
an angle $\alpha$, which depends on the magnetic field and viewing
geometry, on the photon energy and on the position of the point
from which photons were emitted (\S \ref{subsec:calculations}).

The effects induced by rotation of the reference frames compound
those of vacuum polarization.
According to QED, in fact, photons maintain their initial polarization 
state within the adiabatic region, while the polarization freezes at a larger
distance (see \S \ref{subsec:vacuumpolarization}). Despite the transition
between the adiabatic and the outer zones is smooth, we assumed that there is 
a sharp boundary at the adiabatic radius $r_\mathrm{a}$ (see
equation \ref{eqn:ra}). This enabled us to treat the photons as if they
were emitted at $r_\mathrm a$, as far as their polarization state
is concerned. This implies that the distribution of the $\alpha$
angles, by which each frame has to be rotated, is actually
determined by the magnetic field at the adiabatic  radius, and
hence depends also on the distance $r_\mathrm a$ from the star
surface.

Simulations of polarization measurements (\S
\ref{subsec:simulationsresults}) clearly show that, because of the
combined effects of frame rotation and QED, the measured
polarization fraction can be very different from the intrinsic
value, i.e. that of the radiation emitted at the surface. The
differences appear to depend firstly on the viewing geometry, i.e.
on the angles $\chi$ and $\xi$ which give the inclination of the
line-of-sight and of the dipole axis with respect to the star spin
axis. As shown in Figure \ref{fig:plot1}, the polarization
dramatically decreases at all rotational phases for
$\chi\simeq\xi\simeq0$. For $\chi$ not too close to $\xi$,
$\Pi_\mathrm{L}$ has a minimum at the phase $\gamma$ where the
magnetic axis $\boldsymbol{b}_\mathrm{dip}$ lies in the plane of
the rotation axis and the LOS (either $\gamma=0$ or $2\pi$ with
our choice of the reference frame). This behaviour confirms
the results of \citet{Heyletal2003} and it is entirely due
to the non-constant magnetic field across the 
emitting region: when a region of the
star near to the magnetic poles is into view, the projection of
$\boldsymbol{B}$ in the plane orthogonal to the LOS is essentially
radial, so that $\alpha$ can take values in the entire range
$[0,2\pi]$. Since the Stokes parameters for the whole radiation
are obtained integrating the rotated Stokes parameters of single
photons over the part of the star in view (see equations
\ref{eqn:integralStokes}), the polarization degree has a minimum
when the angle between $\boldsymbol{b}_\mathrm{dip}$ and the LOS
is minimum (and equal to $|\chi-\xi|$, see the first of equations \ref{eqn:etazeta}).
On the other hand, the fact that $\Pi_\mathrm{L}$ does not change with rotational
phase in the case shown in the top-left panel of Figure \ref{fig:plot1}, is
precisely due to the fact that this is a nearly aligned rotator viewed
along the rotational axis.

In agreement with the results by \citet{Heyletal2003}, we found
that the behavior of $\Pi_\mathrm{L}$ is also sensitive to the
location of the adiabatic radius. From the top rows of Figures
\ref{fig:plot1} and \ref{fig:plot2}, it can be seen that the
linear polarization fraction increases with the photon energy and
the polar strength of the $B$-field: this reflects the dependence
of $r_\mathrm{a}$ on $E^{1/5}B_\mathrm{P}^{2/5}$. In fact, as
equation (\ref{eqn:cosalpha}) shows, $\cos\alpha$ depends on the
magnetic co-latitude and azimuth, $\theta$ and $\phi$, through
$B_\mathrm X\,, B_\mathrm Y$; the two latter angles contain the
factor $R_\mathrm{NS}/r_\mathrm{a}$ (see equations
\ref{eqn:costhetafull} and \ref{eqn:cosphifull}). So, in the limit
$r_\mathrm{a}\gg R_\mathrm{NS}$ (at least for axisymmetric magnetic field
topology), $\alpha$ remains nearly constant
as the emission point changes on the star surface, implying that
the polarization fraction can be indeed approximated with
$|n_\mathrm{X}- n_\mathrm{O}|/(n_\mathrm{X}+n_\mathrm{O})$, as equation (\ref{eqn:PLunifB})
shows. \citet{Heyletal2003} explained this behavior as due to the fact that, 
in this limit, QED birefringence aligns the photon polarization angles.
Actually, the weaker depolarization when $r_\mathrm{a}\gg R_\mathrm{NS}$ is 
due chiefly to the rotation of the Stokes parameters, vacuum polarization 
entering only implicitly through the dependence of the angle $\alpha$ on 
$r_\mathrm{a}$.
Because of the dependence of $r_\mathrm{a}$ on $E$ and $B_\mathrm{P}$,
this approximation becomes better the larger the photon energy and
the stronger the polar $B$-field. Instead, the closer to the star
surface the adiabatic limit, the smaller the overall measured
polarization degree, the latter becoming vanishingly small if no
adiabatic region is accounted for. So, the main conclusion is the
more point-like the star is seen by an observer at the adiabatic
boundary, the closer the measured $\Pi_\mathrm{L}$ is to the
intrinsic linear polarization degree. A similar effect was noted by \cite{Heyletal2003}
in connection with the variation of the stellar radius.

On the other hand, the polarization angle exhibits quite a
different behaviour. As Figures \ref{fig:plot1} and
\ref{fig:plot2} show, $\chi_\mathrm{p}$ does not change
significantly with $r_\mathrm{a}$, since it does not depend on $E$ and $B_\mathrm P$. This is because
the factors $R_\mathrm{NS}/r_\mathrm{a}$ within $\cos\alpha$ tend to cancel
out taking the ratio $U/Q$ which defines $\chi_\mathrm{p}$ (equation
\ref{eqn:polobs}). The fact that $\chi_\mathrm{p}$ keeps oscillating even when the measured polarization
fraction is much smaller than the intrinsic one (see bottom-left panels
of Figures \ref{fig:plot1} and \ref{fig:plot2}) is a consequence of the frame rotation and not of QED effects.
The polarization angle depends quite strongly, instead, on the geometrical
angles $\chi$ and $\xi$ (see e.g. the right panel of Figure \ref{fig:plotchicsi}).
The polarization swing generally increases
for decreasing $\chi$ at fixed $\xi$, as shown in Figure
\ref{fig:plot1}. In particular, $\chi_\mathrm{p}$ sweeps the
entire range $[0,180^\circ]$ when the region close to the magnetic
pole is always in view during the star rotation (bottom left
panel), while the swing gets smaller for values of $\chi$ and
$\xi$ such that the polar region enters into view only at
certain rotational phases. On the other hand, the oscillation
amplitude in general grows for increasing $\xi$ at fixed $\chi$.
This behavior appears to be related again to the $\alpha$-angle
distribution, and provides an explanation for the correlation between
the swing by $180^\circ$ of the polarization angle and the low phase-averaged
polarization fraction at $\chi < \xi$, as already noticed by
\citet[see also \citealt{WagnerSiefert}]{Fernandez+Davis}. In fact, the regions where the polarization
angle spans the widest range correspond to those in which at least one among the Stokes parameters $Q$
and $U$ takes all the values between $-1$ and $1$. Consequently, the averaged
polarization fraction, obtained by summing the Stokes parameters over a rotational cycle,
turns out to be very small, as shown in Figure \ref{fig:plotchicsi}.

Phase-resolved polarization angle measurements, together with the information
given by the linear polarization fraction, can help in
understanding which polarization mode is the dominant one in the
detected radiation. We showed in Figure \ref{fig:chipolXO} that the mean
value of $\chi_\mathrm{p}$ depends on the mode in which
the majority of photons are polarized. However, it is also related to
the orientation of the $(u,v)$ axes in the polarimeter plane
(the $\psi$ angle, see \S \ref{subsec:calculations}), which are fixed by the instrument design. In
particular, the mean values of $\chi_\mathrm{p}$ for X- and O-mode
photons are always displaced by $90^\circ$, but they are $90^\circ$ and
$0^\circ$ respectively  (as in the case in Figure \ref{fig:chipolXO})
only if $\psi=0$.
Hence, a measurement of the
polarization angle alone fails in telling which is the
prevailing polarization mode. The problem can be solved if also a
phase-resolved measurement of the linear polarization fraction is
available. In this case, since
$\Pi_\mathrm{L}$ has a minimum when
$\boldsymbol{b}_ \mathrm{dip}$ intercepts the
$\boldsymbol{\Omega}-\boldsymbol{\ell}$ plane (see above), it could indeed  be possible to individuate the direction
of the $X$ axis on the plane of the sky. This allows to derive the
angle $\psi$ and to remove
the inherent ambiguity in the measurement of $\chi_\mathrm{p}$.

Polarization observables can also provide information on the
source geometry, i.e. the inclination of the LOS and of the magnetic axis
wrt the rotation axis. In fact, as discussed earlier on, both the
polarization fraction and  the polarization angle strongly depend
on the angles $\chi$ and $\xi$. As already shown in
\citet{Tavernaetal}, if phase-resolved polarization signals are
available, a simultaneous fit of $\Pi_\mathrm{L}$ and
$\chi_\mathrm{p}$ (possibly supplemented by that of the flux)
allows to unequivocally derive the values of $\chi$ and $\xi$. On
the contrary, this is in general not possible starting from
phase-averaged measurements. The phase-averaged polarization
fraction is largely degenerate with respect to the two angles, as
clearly shown in Figure \ref{fig:plotchicsi} 
and, since the phase average polarization angle is constant
in large regions of the $\chi-\xi$ plane, its measure is of
no avail in pinpointing $\chi$ and $\xi$.

The effects of a different magnetic field topology on the
polarization observables were assessed in the illustrative case of
globally-twisted dipolar magnetic field \footnote{We focused here
only on surface emission, the interactions of photons with
magnetospheric currents, chiefly through resonant cyclotron
scattering, were ignored.}. The presence of a toroidal component
in the external magnetic field slightly changes the behaviour of
linear polarization fraction (see Figure \ref{fig:plot4}). In a
twisted field, depolarization induced by the frame rotation is a
bit stronger. This is due to the fact that $B_{\mathrm{twist}}>
B_{\mathrm{dip}}$ at any given position, because the toroidal
component is roughly of the same order of the poloidal one, while
the $r$-dependence is about the same for the two magnetic
configurations. As a consequence the adiabatic boundary moves a
bit closer to the surface if  $B_\mathrm{P}$ and the photon energy
are the same. A twisted field influences the polarization angle
most, producing a strong asymmetry in the swing and a weak dependence
on the energy (mainly at optical energies), as already discussed by \citet{Fernandez+Davis}
and \citet{Tavernaetal}. The fact that
the polarization angle is more sensitive to QED effects for a twisted magnetosphere
than for a purely dipolar field, provides a strong signature of vacuum
polarization effects \cite[see][]{Tavernaetal}.

Our analysis further demonstrates the need to properly account 
for QED and frame rotation effects in evaluating the observed polarization properties of
radiation emitted by a neutron star. This is of particular relevance in relation 
to recently proposed X-ray polarimetry missions, which will certainly select 
neutron star sources as primary targets. 

\section*{Acknowledgments}
It is a pleasure to thank Enrico Costa for many illuminating
discussions, Kinwah Wu for some useful comments and an anonymous referee, 
whose helpful suggestions helped us in improving a previous version of 
the manuscript. The work of RT is partially supported by INAF through a PRIN grant.
DGC aknowledges a fellowship from CONICYT-Chile (Becas Chile). He
also aknowledges financial support from the RAS and the University
of Padova for funding a visit to the Department of Physics and
Astronomy, during which part of this investigation was carried
out.

\appendix{} \label{sec:appendix}

\section{Cartesian components of \textit{B}} \label{subsec:Bfield}
The cartesian compononents of the magnetic field $\boldsymbol{B}$ in the
reference frame $(p,q,t)$ can be obtained from its polar
components $\boldsymbol{B}^\mathrm{pol}=(B_r,B_\theta,B_\phi)$ given in
equation (\ref{eqn:Bdippolar}) using the following expression
\begin{flalign} \label{eqn:Bdipcartesdef}
\boldsymbol{B} &= (B_p,B_q,B_t) = (\boldsymbol{B}^\mathrm{pol}\cdot
\boldsymbol{p}^\mathrm{pol},\boldsymbol{B}^\mathrm{pol}\cdot
\boldsymbol{q}^\mathrm{pol},\boldsymbol{B}^\mathrm{pol}\cdot
\boldsymbol{t}^\mathrm{pol})\,, &
\end{flalign}
where $\boldsymbol{p}^\mathrm{pol}$, $\boldsymbol{q}^\mathrm{pol}$ and
$\boldsymbol{t}^\mathrm{pol}$ are the unit vectors relative to the
$(p,q,t)$ frame expressed in polar components
\begin{flalign} \label{eqn:pqrpolar}
\boldsymbol{p}^\mathrm{pol} &= \boldsymbol{p}\cdot(\hat{r},\hat{\theta},
\hat{\phi}) = (\sin\theta\cos\phi,\cos\theta\cos\phi,-\sin\phi) &
\notag \\
\boldsymbol{q}^\mathrm{pol} &= \boldsymbol{q}\cdot(\hat{r},\hat{\theta},
\hat{\phi}) = (\sin\theta\sin\phi,\cos\theta\sin\phi,\cos\phi) & \\
\boldsymbol{t}^\mathrm{pol} &= \boldsymbol{t}\cdot(\hat{r},\hat{\theta},
\hat{\phi}) = (\cos\theta,-\sin\theta,0)\,, & \notag
\end{flalign}
and the angles $\theta$ and $\phi$ are the magnetic colatitude and
azimuth, respectively (see Figure \ref{fig:LOSbdipframe}b).
Upon substituting expressions (\ref{eqn:pqrpolar}) into equation
(\ref{eqn:Bdipcartesdef}), one finally  obtains
\begin{flalign} \label{eqn:Bdipcartes}
B_p &= \sin\theta\cos\phi B_r + \cos\theta\cos\phi B_\theta - \sin\phi
B_\phi & \notag \\
B_q &= \sin\theta\sin\phi B_r + \cos\theta\sin\phi B_\theta + \cos\phi
B_\phi & \\
B_t &= \cos\theta B_r - \sin\theta B_\theta\,. & \notag
\end{flalign}

\section{Magnetic reference frame} \label{subsec:pqtframe}
The projection of $\boldsymbol{b}_\mathrm{dip}$, given by equation
(\ref{eqn:bdip}), orthogonal to the spin axis $\boldsymbol{\Omega}$
(see \S \ref{subsec:calculations}), in the LOS reference frame $(X,Y,Z)$ is
\begin{flalign} \label{eqn:hatq}
\boldsymbol{m} &\equiv \frac{\boldsymbol{b}_\mathrm{dip}-
(\boldsymbol{b}_\mathrm{dip}\cdot\boldsymbol{\Omega})
\boldsymbol{\Omega}}{|\boldsymbol{b}_\mathrm{dip}-
(\boldsymbol{b}_\mathrm{dip}\cdot\boldsymbol{\Omega})
\boldsymbol{\Omega}|} = \left(\begin{array}{c}
-\cos\chi\cos\gamma \\
\sin\gamma \\
\sin\chi\cos\gamma
\end{array}\right)\,; &
\end{flalign}
$\boldsymbol{m}$ is an unit vector corotating with the star
around the spin axis. The projection of $\boldsymbol{m}$ perpendicular to
$\boldsymbol{b}_\mathrm{dip}$ fixes the $p$ axis of the
$\boldsymbol{b}_\mathrm{dip}$ reference frame $(p,q,t)$. Its expression in the $(X,Y,Z)$ frame is
given by
\begin{flalign} \label{eqn:pXYZ}
\boldsymbol{p} &\equiv \frac{\boldsymbol{m}-(\boldsymbol{m}\cdot
\boldsymbol{b}_\mathrm{dip})\boldsymbol{b}_\mathrm{dip}}{|\boldsymbol{m}-
(\boldsymbol{m}\cdot\boldsymbol{b}_\mathrm{dip})
\boldsymbol{b}_\mathrm{dip}|} & \notag \\
\ &= \left(\begin{array}{c}
-\sin\chi\sin\xi-\cos\chi\cos\xi\cos\gamma \\
\cos\xi\sin\gamma \\
\sin\chi\cos\xi\cos\gamma-\cos\chi\sin\xi
\end{array}\right)\,. &
\end{flalign}
Finally, the unit vector defining the $q$ axis, in
the $(X,Y,Z)$ reference frame, is given by the vector product
between $\boldsymbol{b}_\mathrm{dip}$ and $\boldsymbol{p}$
\begin{flalign} \label{eqn:qXYZ}
\boldsymbol{q} &= \boldsymbol{b}_{dip}\times\boldsymbol{p} =
\left(\begin{array}{c}
-\cos\chi\sin\gamma \\
-\cos\gamma \\
\sin\chi\sin\gamma
\end{array}\right)\,. &
\end{flalign}

\section{Complete expressions for $\cos\theta$ and $\cos\phi$}
\label{subsec:costhetacosphi}
Substituting the components of
$\boldsymbol{b}_\mathrm{dip}$ (equation \ref{eqn:bdip}) and of
$\boldsymbol{r}_\mathrm{a}$ (equation \ref{eqn:ravector}) into equation (\ref{eqn:costheta}) one obtains
\begin{flalign} \label{eqn:costhetafull}
\cos\theta &= \frac{R_\mathrm{NS}}{r_\mathrm{a}}\sin\Theta_\mathrm{S}\bigg(\cos\Phi_\mathrm{S}\sin\chi\cos\xi +
\sin\Phi_\mathrm{S}\sin\xi\sin\gamma & \notag \\
\ &\ \ \ \ \ \ \ \ \ \ \ \ \ \ \ \ \ \ \ \ \ \ \ \ \ \ \  -\cos\Phi_\mathrm{S}\cos\chi\sin\xi\cos\gamma\bigg) & \notag \\
\ &+\sqrt{1-\left(\frac{R_\mathrm{NS}}{r_\mathrm{a}}\sin\Theta_\mathrm{S}\right)^2}
\bigg(\cos\chi\cos\xi + \sin\chi\sin\xi\cos\gamma\bigg)\,. &
\end{flalign}
To calculate the complete expression of $\cos\phi$ we use equation
(\ref{eqn:cosphi}), where the components of the unit vector
$\boldsymbol{p}$ are given in equation (\ref{eqn:pXYZ}) and
\begin{flalign} \label{eqn:raperp}
\boldsymbol{r}_\mathrm{a}^\perp &= \frac{\boldsymbol{r}_\mathrm{a} -
(\boldsymbol{r}_\mathrm{a}\cdot\boldsymbol{b}_\mathrm{dip})
\boldsymbol{b}_\mathrm{dip}}{|\boldsymbol{r}_\mathrm{a} -
(\boldsymbol{r}_\mathrm{a}\cdot\boldsymbol{b}_\mathrm{dip})
\boldsymbol{b}_\mathrm{dip}|}\,. &
\end{flalign}
Putting all together, one obtains
\begin{flalign} \label{eqn:cosphifull}
\cos\phi &= \frac{R_\mathrm{NS}\sin\Theta_\mathrm{S}}{r_\mathrm{a}\sin\theta}
\bigg(\sin\Phi_\mathrm{S}\cos\xi\sin\gamma-\cos\Phi_\mathrm{S}\sin\chi\sin\xi & \notag \\
\ &\ \ \ \ \ \ \ \ \ \ \ \ \ \ \ \ \ \ \ \ \ \ \ \ \ \ \ -\cos\Phi_\mathrm{S}\cos\chi\cos\xi\cos\gamma\bigg) & \notag \\
\ &+ \sqrt{\frac{r_\mathrm{a}^2-(R_\mathrm{NS}\sin\Theta_\mathrm{S})^2}{r_\mathrm{a}^2\sin^2\theta}}
\bigg(\sin\chi\cos\xi\cos\gamma-\cos\chi\sin\xi\bigg)\,. &
\end{flalign}

\label{lastpage}

\end{document}